\documentclass[10pt,epsfig]{article}
\usepackage{graphicx}
\usepackage{array}
\usepackage{amsmath,amssymb}
\usepackage[latin1]{inputenc}

\begin{document}


\title{The $Z_3$-graded extension of the Poincar\'e algebra}

\author{Richard Kerner}

\date{}

\maketitle

{\begin{center}
{\it Laboratoire de Physique Th\'eorique de la Mati\`ere Condens\'ee (LPTMC),}
\vskip 0.1cm
{\it Sorbonne - Universit\'es - CNRS UMR 7600} 
\vskip 0.1cm
{\it Tour 23-13, 5-\`eme \'etage, Bo\^{i}te Courrier 121, 4 Place Jussieu, 75005 Paris, FRANCE} 
\end{center}}

\vskip 0.5cm

\begin{abstract}
A $Z_3$ symmetric generalization of the Dirac equation was proposed in \cite{Kerner2017}, \cite{Kerner2018}, and its
properties and solutions discussed in \cite{Kerner2018B}, \cite{Kerner2019}. The generalized Dirac operator acts on
{\it coloured spinors} composed out of six Pauli spinors, describing three colours and particle-antiparticle degrees
 of freedom characterizing a single quark state, thus combining $Z_2 \times Z_2 \times Z_3$ symmetries of $12$-component
generalized wave functions. The $Z_3$-graded generalized Lorentz algebra and its spinorial representation were introduced
in \cite{RKJL2019}, leading to the appearance of extra $Z_2 \times Z_2 \times Z_3$ symmetries, probably englobing 
the symmetries of isospin, flavors and families.

The present article proposes the construction of $Z_3$-graded extension of the Poincar\'e group. It turns out that 
such a generalization requires introduction of extended $12$-dimensional Minkowskian space-time containing the usual
$4$-dimensional space-time as a subspace, and two other mutually conjugate ``replicas" with complex-valued vectors and metric
tensors. Representation in terms of differential operators and generalized Casimir operators are introduced and their
symmetry properties are briefly discussed. 

\end{abstract}

\section{Introduction}

There can be little doubt that of all symmetries displayed by physics of elementary particles and fields, 
the invariance under the action of discrete groups are by far the best confirmed by the experiment, and also the most fundamental. 
The simplest discrete group is $S_2$, the group of permutations of two objects. These permutations are cyclic, therefore
 the same group can be interpreted as $Z_2$. Such an identification is no more possible for the next permutation group,
 $S_3$, which contains six elements, out of which only the cyclic ones form a three-dimensional subgroup $Z_3$. 

The cyclic group $Z_2$ plays crucial role in quantum physics of particles and fields. Its two representations in complex plane,
implemented as symmetries of arguments of complex wave functions, the trivial one and the faithful one, lead to different 
fundamental statictics creating the great divide between two quantum statistics characterizing bosons and fermions.
In the space of functions depending on two arguments, we can have two dfferent behaviors with repsect to the permutations;
\begin{equation}
\Psi(x, y) = \Psi(y,x) \; \; {\rm for \; bosons, \; \; \; and} \; \;  \Psi (x, y) = - \Psi (y, x) \; \; {\rm for \; fermions}.
\label{bosefermi}
\end{equation}
The relationship between the spin and statistics is another illustration of importance of discrete symmetries. It establishes
a one-to-one dependence between the ireeducible representations of the Lorentz group and the two possible statistical
behaviors defined by (\ref{bosefermi}): half-integer spin representations for fermions, and integer spin representations for
bosons.     

The Dirac equation for the electron provides and example of entaglement of two apparently independent $Z_2$ symmetries.
The discovery of dichotomic spin parameter which in the case of the electron can take on only two exclusive values led
Pauli to the conclusion that a Sch\"odinger-like equation for the electron should involve a two-component wave function:
\begin{equation} 
E \begin{pmatrix} \psi^1 \cr \psi^2 \end{pmatrix} =
mc^2 \;  \begin{pmatrix} \psi^1 \cr \psi^2 \end{pmatrix} + c \;{\boldsymbol{\sigma}} \cdot {\bf p} \begin{pmatrix} \psi^1 \cr \psi^2 \end{pmatrix}
\label{Paulihalf}
\end{equation}
where ${\boldsymbol{\sigma}} = (\sigma_x, \; \sigma_y, \; \sigma_z)$ denotes the three Pauli matrices, which form the basis of
$2 \times 2$ traceless hermitian matrices, $E = - i \hbar \partial_t$ and ${\bf p} = (p^x, \; p^y, \; p^z)$ is the momentum operator,
 with $p_k = - i \hbar \partial_k$.
This equation does not satisfy the Lorentz-invariant condition: iterating it leads to wrong relation between energy, momentum and mass,
$E^2 = m^2 c^4 + 2 m c {\bf p} + {\bf p}^2$ instead of $E^2 = m^2 c^4 + c^2 {\bf p}^2$, which made Pauli abandon this version
introducing the approximate non-relativistic equation for the electron interacting with electromagnetic field \cite{Pauli1926}.

It turns out that relativistic covariance can be restored via introduction of another two-component Pauli spinor, the two similar
equation intertwining them with mass terms of opposite sign. Let the two Pauli spinors be denoted by $\psi_{+}$ and $\psi_{-}$. Then
the following system of equations satisfies relativistic dispersion relation, and is Lorentz covariant:
\begin{equation}
E \psi_{+} = mc^2 \psi_{+} + {\boldsymbol{\sigma}} \cdot {\bf p} \psi_{-},  \; \; \; \; \; 
E \psi_{-} = - mc^2 \psi_{-} + {\boldsymbol{\sigma}} \cdot {\bf p} \psi_{+},
\label{Dirac1}
\end{equation}
which is Dirac's equation in a less usual basis.\cite{Dirac1928}. By iterating it, we get the relativistic condition satisfied
bu both Pauli spinors:
$ E^2 \psi_{+} = (m^2 c^4 + c^2 {\bf p}^2 ) \psi_{+}, \; \; \; \; E^2 \psi_{-} = (m^2 c^4 + c^2 {\bf p}^2 ) \psi_{-} $.
In the more familiar form, the same system is written in a manifestly relativistic form, with the $4$-component Dirac spinor composed of two Pauli 
spinors $\psi_{+}$ and $\psi_{-}$, and the $4 \times 4$ Dirac matrices expressed in terms of tensor products of $2 \times 2$ matrices as follows:
\begin{equation}
\gamma^{\mu} p_{\mu} \; \psi = m \psi, \; \; \; {\rm where} \; \; \; 
\gamma^0 = \sigma_3 \otimes {\mbox{l\hspace{-0.55em}1}}_{2}, \; \; \; \; \gamma^i = (i \sigma_2) \otimes \sigma^i.
\label{Diracpure}
\end{equation}
The Dirac equation is invariant with respect to $Z_2 \times Z_2$ symmetry. The first $Z_2$ concerns the spin of the electron,
which can have two projections on the momentum; the second $Z_2$ group, imposed by the requirement of Lorentz invariance,
concerns the particle-antiparticle symmetry.

Recently in \cite{RKOS2014}, \cite{Kerner2017}, \cite{Kerner2018}, \cite{Kerner2018B} a generalization of the Dirac equation for quarks
was proposed, incorporating the color degrees of freedom via extending the discrete symmetry of the system to the  $Z_2 \times Z_2 \times Z_3$
group. The cyclic group $Z_3$ is generated by the third root of unity, denoted by $j = e^{\frac{2 \pi i}{3}}$, with $j^2 =   e^{\frac{4 \pi i}{3}}$,
$j^3 =1$, and $1+j+j^2 = 0$. Just as taking into account the dichotomic half-integer spin variable, the introduction of color degrees
of freedom requires additional $Z_3$ symmetry acting on a new discrete variable taking three possible (and exclusive) values, named symbolically
``red", ``blue" and ``green". The $3 \times 3$ matrices had to be introduced, all representing {\it third roots} of the $3 \times 3$ unit matrix.
 Six Pauli spinors represent three colors and three anti-colors:
{\small
\begin{equation} 
\varphi_{+} = \begin{pmatrix} \varphi_{+}^1 \cr \varphi_{+}^2 \end{pmatrix}, \; \; 
\chi_{+} = \begin{pmatrix} \chi_{+}^1 \cr \chi_{+}^2 \end{pmatrix}, \; \; \; 
\psi_{+} = \begin{pmatrix} \psi_{+}^1 \cr \psi_{+}^2 \end{pmatrix}, \; \; 
\varphi_{-} = \begin{pmatrix} \varphi_{-}^1 \cr \varphi_{-}^2 \end{pmatrix}, \; \;  
\chi_{-} = \begin{pmatrix} \chi_{-}^1 \cr \chi_{-}^2 \end{pmatrix}, \; \; 
\psi_{-} = \begin{pmatrix} \psi_{-}^1 \cr \psi_{-}^2 \end{pmatrix},
\label{sixPauli}
\end{equation}}
on which Pauli sigma-matrices act in a natural way. By analogy with the pair of equations (\ref{Dirac1}) in which multiplying the mass by $-1$
led to the  anti-particle appearance, now the mass term is multipleid by the generator of the $Z_3$ group, $j$, each time the colour changes.
This yields the following set of what may be called the "colour Dirac equation":
$$E \; \varphi_{+} = mc^2 \, \varphi_{+} + c \; {\boldsymbol{\sigma}} \cdot {\bf p} \, \chi_{-}, \; \; \; \; \; 
E \; \varphi_{-} = - mc^2 \, \varphi_{-} + c \; {\boldsymbol{\sigma}} \cdot {\bf p} \, \chi_{+}$$
$$E \; \chi_{+} = j \; mc^2 \, \chi_{+} + c \; {\boldsymbol{\sigma}} \cdot {\bf p} \, \psi_{-} ,\; \; \; \; \; 
E \; \chi_{-} = - j \; mc^2 \, \chi_{-} + c \; {\boldsymbol{\sigma}} \cdot {\bf p} \, \psi_{+}$$
\begin{equation}
E \; \psi_{+} = j^2 \;  mc^2 \, \psi_{+} + c \; {\boldsymbol{\sigma}} \cdot {\bf p} \, \varphi_{-}, \; \; \; \; \; 
E \; \psi_{-} = -j^2 \;mc^2 \, \psi_{-} + c \; {\boldsymbol{\sigma}} \cdot {\bf p} \, \varphi_{+}
\label{systemsix}
\end{equation}
In an appropriate basis, the system (\ref{systemsix}) can be represented in a Dirac-like form as follows:
\begin{equation}
\Gamma^{\mu} p_{\mu} \Psi = mc \; {\mbox{l\hspace{-0.55em}1}}_{12} \; \Psi,
\label{Dirac12}
\end{equation} 
where $\Psi$ is the generalized $12$-component spinor made of$6$ Pauli spinors (\ref{sixPauli}), and the generalized
$12 \times 12$ Dirac matrices $\Gamma^{\mu}$ are constructed as follows:
\begin{equation}
\Gamma^0 = B^{\dagger} \otimes \sigma_3 \otimes {\mbox{l\hspace{-0.55em}1}}_{2}, \; \; \; 
\Gamma^i = Q_2 \otimes (i \sigma_2) \otimes \sigma^i,
\label{colourDmat}
\end{equation}
where
\begin{equation}
B^{\dagger} = \begin{pmatrix} 1 & 0 & 0 \cr 0 & j^2 & 0 \cr 0 & 0 & j \end{pmatrix}, \; \; \; \; 
B = \begin{pmatrix} 1 & 0 & 0 \cr 0 & j & 0 \cr 0 & 0 & j^2 \end{pmatrix}, \; \; \; \; 
Q_2 = \begin{pmatrix} 0 & 1 & 0 \cr 0 & 0 & j^2  \cr j & 0 & 0 \end{pmatrix}, 
\label{BQmatrix}
\end{equation}
The two traceless matrices $B$ and $Q_2$ are both cubic roots of unit $3 \times 3$ matrix. They generate
the entire Lie algebra of the $SU(3)$ group. 

The system (\ref{Dirac12}) becomes diagonal only after sixth iteration, yielding the dispersion relation of sixth order:
\begin{equation}
( \Gamma^{\mu} p_{\mu} )^6 = (p_0^6 - {\bf p }^6) \;  {\mbox{l\hspace{-0.55em}1}}_{2} = m^6 c^6 \; {\mbox{l\hspace{-0.55em}1}}_{2}
\label{dispersix}
\end{equation}
This expression is not manifestly relativistic invariant, but it represents a unique light cone multiplied by a positive form-factor:
\begin{equation}
(p_0^6 - {\bf p }^6) = (p_0^2 - {\bf p}^2)(j p_0^2 - {\bf p}^2)(j^2 p_0^2 - {\bf p}^2) = 
(p_0^2 - {\bf p}^2)(p_0^4 + p_0^2 {\bf p}^2 + \mid {\bf p} \mid^4.
\label{pcone}
\end{equation}
Such field theories of higher order were considered by T.D.Lee and G.Wick \cite{LeeWick} and were recently an object of renewed
interest \cite{AnselmiPiva}.

The colour Dirac matrices $\Gamma^{\mu}$ defined in (\ref{colourDmat}) do not span the usual Clifford algebra, and do not
transform as relativistic $4$-vectors under ordinary Lorentz transformations. In order to implement Lorentz covariance, the
set of $\Gamma$-matrices must be extended up to six different realizations, forming doublets transforming under the
extension of the Lorentz algebra, containing the usual Lorentz algebra as subalgebra, and two conjugate replicas
forming a $Z_3$-graded algebra 
${\cal{L}} = {\overset{(0)}{\cal{L}}}\oplus {\overset{(1)}{\cal{L}}}\oplus {\overset{(2)}{\cal{L}}},$
acting on the generalized multi-spinors formed by six $12$-dimensional colour Dirac spinors, the total dimension
of the representation space being $6 \times 12 = 72$ (see the details in \cite{RKJL2019}. The multiplication
rules in $L$ are $Z_3$-graded, i.e. one has ${\overset{(r)}{\cal{L}}} \cdot {\overset{(s)}{\cal{L}}} \subset {\overset{((r+s)\mid_3)}{\cal{L}}}$.

The aim of the present article is to define a similar $Z_3$-graded extension of the Poincar\'e algebra realized in terms
of differential operators acting on an extended Minkowskian space-time.

\section{The $Z_3 \times Z_2$ symmetry}

Let us recall briefly the properties of the cyclic ($Z_3$) and the permutation ($S_3$) groups of three elements. 
Their representation in terms of rotations and reflections in the complex plane are shown in the following Figure
\ref{fig:Rotations}:
\begin{figure}[hbt]
\centering 
\includegraphics[width=3.6cm, height=3.7cm]{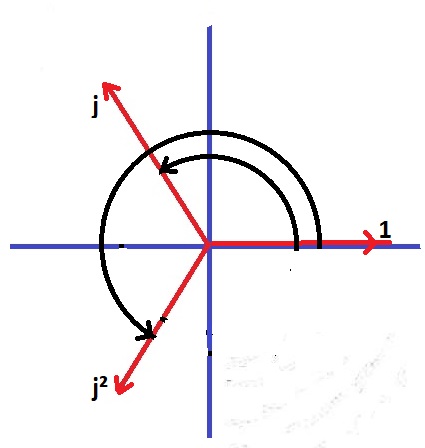} 
\hskip 0.5cm
\includegraphics[width=3.6cm, height=3.7cm]{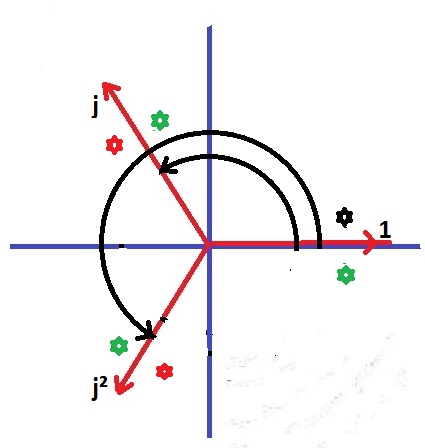} 
\caption{ Rotations ($Z_3$-group) and reflections added ($S_3$ group)  }
\label{fig:Rotations}
\end{figure}
Let us denote by $j$ and $j^2$ the two complex third roots of unity, given by
\begin{equation}
j = e^{\frac{2 \pi i}{3}} = - \frac{1}{2} + \frac{i \sqrt{3}}{2}, \; \; \; j^2 = e^{\frac{4 \pi i}{3}} = - \frac{1}{2} - \frac{i \sqrt{3}}{2}
\label{jj2}
\end{equation}
satisfying obvious identities $1+j+j^2 = 0,$  so that $ j+j^2 = -1, \; \;  j-j^2 = i \sqrt{3}$,

The six $S_3$ symmetry transformations contain the identity, two rotations, one by  $120^o$, another one  by $240^o$, and three
reflections, in the $x$-axis, in the $j$-axis and in the $j^2$-axis. 
The $Z_3$ subgroup contains only the three rotations. Odd permutations must be represented by idempotents, i.e. by operations 
whose square is the identity operation. We can make the following choice:
{\small
\begin{equation}
\begin{pmatrix}
ABC \cr CBA
\end{pmatrix}
\rightarrow ({\bf z \rightarrow {\bar{z}}}), \, \ \ \, 
\begin{pmatrix}
ABC \cr BAC
\end{pmatrix}
\rightarrow ({\bf z \rightarrow {\hat{z}}}), \, \ \ \,
\begin{pmatrix}
ABC \cr CBA
\end{pmatrix}
\rightarrow ({\bf z \rightarrow z^{*}}),
\label{permutationsodd}
\end{equation}}
Here the bar $({\bf z \rightarrow {\bar{z}}})$ denotes the complex conjugation, i.e. the reflection
in the real line, the hat  ${\bf z \rightarrow {\hat{z}}}$ denotes the reflection in the root $j^2$,
and the star ${\bf z \rightarrow z^{*}}$ the reflection in the root $j$.
The six operations close in a non-abelian group with six elements, which are represented as rotation and reflexion
operation in the complex plane, as shown in (\ref{fig:Rotations}) above. 

In what follows, we shall use the $Z_3$ group for grading of linear spaces and matrix algebras
\cite{Kerner1991}, \cite{AbrKer1997}, \cite{RKOS2014}. The $Z_3$-graded algebras are composed of
three vector subspaces, one of which (of $Z_3$-grade zero) constitutes a subalgebra in the ordinary sense:
\begin{equation}
{\cal{A}}= {\cal{A}}_0 \oplus {\cal{A}}_1 \oplus {\cal{A}}_2
\label{AZ_3}
\end{equation} 
The multiplication in the graded algebra (\ref{AZ_3}) obeys the following scheme:
\begin{equation}
{\overset{(r) }{\cal{A}}}\cdot  {\overset{(s) }{\cal{A}}} \subset {\overset{(r+s) \mid_3 }{\cal{A}}}, 
\; \; {\rm with} \; \; r, s,.. = 0,1,2, \; \; 
(r+s) \mid_3 = (r+s) \; {\rm modulo} \; \; 3.
\label{Athreegrade}
\end{equation}

The $Z_3$ symmetry can be combined with the $Z_2$ symmetry; $3$ and $2$ being prime numbers, the Cartesian product
of the two is isomorphoic with another cyclic group, $Z_3 \times Z_2 = Z_6$. The generalized Dirac equation is invariant
under the discrete group $Z_3 \times Z_2 \times Z_2 \simeq Z_6 \times Z_2$ (which is not isomorphic with $Z_{12}$  
because $6$ is not a prime number, being divisible by $2$ and by $3$). 

The cyclic group $Z_6$ is represented in the complex plane by its generator $q=e^{\frac{2 \pi i}{6}} = e^{\frac{\pi i}{3}}$,
and its powers from $1$ to $6$. In terms of the $Z_3$ group generated by $j$ and $Z_2$ group generated by $-1$, we have 
$$q = - j^2, \; \; q^2 = j, \; q^3 = -1, \; q^4 = j^2, \; q^5 = - j, \; q^6 = 1,$$
as shown in the figure (\ref{fig:CyclicZ6}) below.
\begin{figure}[hbt]!!
\centering
\includegraphics[width=5.3cm, height=5.3cm]{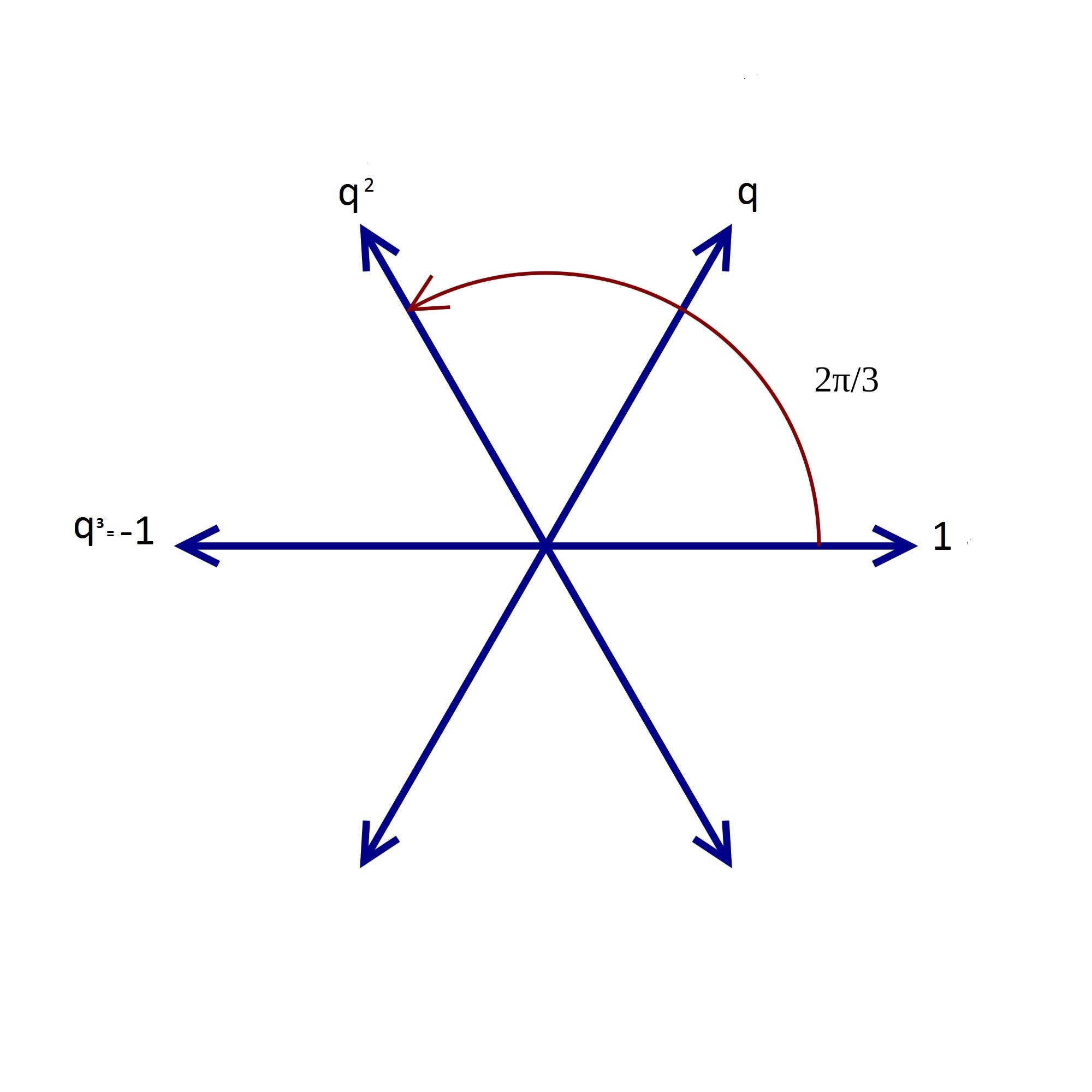}
\hskip 0.4cm
\includegraphics[width=4.7cm, height=4.7cm]{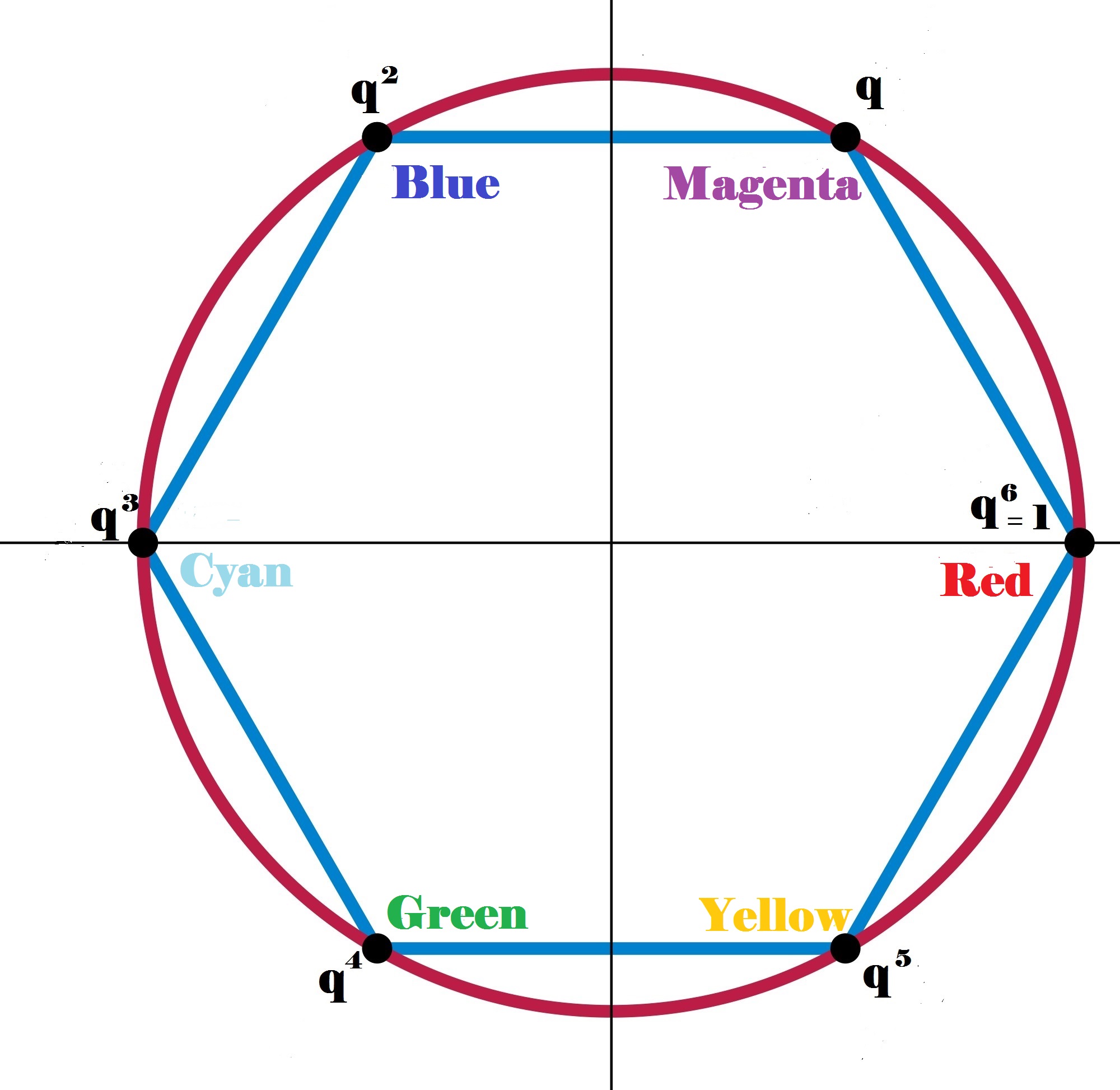}
\caption{\small{The six complex numbers $q^k$ can be put into correspondence with three colours and three anti-colours.}}
\label{fig:CyclicZ6}
\end{figure}
In analogy with colours labeling quark fields, if the ``white" combination is represented by $0$, then we
have {\it two} linear colourless sums of three powers of $q$, namely $1 + q^2 + q^4 =0$ and $q + q^3 + q^5 =0$,
and {\it three} white combinations of colour with its anti-colour, $q + q^4=0, \; q^2 + q^5 = 0, \; q^3 + q^6 =0$,
just like a fermion and its antiparticle, or three bosons (like e.g. mesons $\pi^0, \; \pi^{+}$ and $\pi^{-}$).   

A $Z_3$-graded analog of Pauli's exclusion principle was introduced and its algebraic and physical consequences investigated
in \cite{Kerner2017}, \cite{Kerner2019}.

\section{The $Z_3$-extended Minkowskian spacetime}

Let us denote by $M_4$ the standard four-dimensional Minkowskian spacetime, a $4$-dimensional real vector space
endpwed with pseudo-Euclidean (Minkowskian) metric $\eta_{\mu \nu} = {\rm diag} [+,-,-,-]$. A spacetime vector
is given by its coordinates in a chosen orthonormal frame: 
\begin{equation}
k^{\mu} = [k^0, {\bf k} ] = [ k^0 , k^x, k^y, k^z ]
\label{Kfirst}
\end{equation}
often replaced by a more practical notation with small Greek indices running from $0$ to $3$:
 \begin{equation}
k^{\mu} = [k^0, {\bf k} ] = [ k^0 , k^1, k^2, k^3 ]
\label{Ksecond}
\end{equation}
The three replicas of a $4$-vector $k^{\mu}$ will be labeled with the superscripts relative to the 
elements of the $Z_3$-group as follows:
\begin{equation} 
{\overset{(0) \; \; }{k^{\mu}}} = ( {\overset{(0) \; \; }{k^{0}}}, \; {\overset{(0) \; \; }{\bf{k}}} ), \; \; \; \; \; 
{\overset{(1) \; \; }{k^{\mu}}} = ( {\overset{(1) \; \; }{k^{0}}}, \; {\overset{(1) \; \; }{\bf{k}}} ), \; \; \; \; \;
{\overset{(2) \; \; }{k^{\mu}}} = ( {\overset{(2) \; \; }{k^{0}}}, \; {\overset{(2) \; \; }{\bf{k}}} ).
\label{threekays}
\end{equation}
In each of the three sectors the specific quadratic form is given, defining the group of transformations keeping it invariant:
\begin{equation}
({\overset{(0)}{k^{0}}})^2 - ( {\overset{(0)}{{\bf{k}}}})^2 = m^2, \; \; \; \;
({\overset{(1)}{k^{0}}})^2 - j \; ( {\overset{(1)}{{\bf{k}}}})^2 = j \; m^2, \; \; \; \;
({\overset{(2)}{k^{0}}})^2 - j^2 \; ( {\overset{(2)}{{\bf{k}}}})^2 = j^2 \;m^2,
\label{threeinvs}
\end{equation}
which leads to the following explicit expressions of $({\overset{(r) \; }{k^{0}}})$ as functions of ${\overset{(r)}{\bf{k}}}$ and $m$ ($r=0,1,2$):
\begin{equation}
{\overset{(0) \; }{k^{0}}} = \pm \root \of{ {\overset{(0)}{{\bf{k}}^2}} + m^2}, \; \; \; \; \;
{\overset{(1) \; }{k^{0}}} = \pm j \; \root \of{ {\overset{(1)}{{\bf{k}}^2}} + m^2}, \; \; \; \; \;
{\overset{(2) \; }{k^{0}}} = \pm j^2 \; \root \of{ {\overset{(2)}{{\bf{k}}^2}} + m^2}, 
\label{Threeekzero}
\end{equation}  
Let us denote the three quadratic forms, one real and two mutually complex conjugate, by the following three tensors
\begin{equation}
{\overset{(0)}{\eta}}_{\mu \nu} = {\rm diag} [+1, -1, -1, -1], \; \; \; \; 
{\overset{(1)}{\eta}}_{\mu \nu} = {\rm diag} [+1, -j, -j, -j], \; \; \; \; 
{\overset{(2)}{\eta}}_{\mu \nu} = {\rm diag} [+1, -j^2, -j^2, -j^2] 
\label{threeetas}
\end{equation}
defined on each of the subspaces of the generalized Minkowskian space 
\begin{equation}
{\overset{(Z_3)}{M}}_{12} = {\overset{(0)}{M}}_{4} \oplus {\overset{(1)}{M}}_{4} \oplus {\overset{(2)}{M}}_{4}
\label{Threemink}
\end{equation}
The superscripts $(r) = (0), \; (1), \; (2)$ refer to the $Z_3$-grades attributed to each of the three subspaces.
These grades will play an important role in defining the $Z_3$-graded extension of the Poincar\'e algebra acting
on the extended Minkowskian space-time ${\overset{(Z_3)}{M}}_{12}$.
We should underline here that the three ``replicas" are to be treated as really independent components of the resulting
$12$-dimensional manifold. For convenience, we shall use the same letters designing three types of space-time components, labeling
them with an extra index as follows:
\begin{equation}
x^{\mu}_r = (x^{\mu}_0, \; x^{\mu}_1, \; x^{\mu}_2) = [ \tau_0, x_0, y_0, z_0; \; \tau_1, x_1, y_1, z_0; \; \tau_2, x_2, y_2, z_2 ].
\label{Z3coordinates}
\end{equation}
Idempotent operators projecting on one of the three subspaces of the generalized Minkowskian space-time  ${\overset{(Z_3)}{M}}_{12}$
can be constructed using the $3 \times 3$ matrices $B$ and $B^{\dagger}$ introduced in (\ref{BQmatrix}) as follows. Let us define
two $12 \times 12$ matrices acting on ${\overset{(Z_3)}{M}}_{12}$:
$${\cal{B}} = B \otimes {\mbox{l\hspace{-0.55em}1}}_{4}, \; \; \; {\cal{B}}^{\dagger} = B^{\dagger} \otimes {\mbox{l\hspace{-0.55em}1}}_{4},$$ 
Then the following three projection operators can be formed:
\begin{equation}
{\overset{(0)}{\Pi}} = \frac{1}{3} \; (  {\mbox{l\hspace{-0.55em}1}}_{12} + {\cal{B}} + {\cal{B}}^{\dagger}), \; \; \;
{\overset{(1)}{\Pi}} = \frac{1}{3} \; (  {\mbox{l\hspace{-0.55em}1}}_{12} + j^2 \; {\cal{B}} + j \; {\cal{B}}^{\dagger}), \; \; \; 
{\overset{(2)}{\Pi}} = \frac{1}{3} \; (  {\mbox{l\hspace{-0.55em}1}}_{12} + j \; {\cal{B}} + j^2 \; {\cal{B}}^{\dagger},
\label{threeproj}
\end{equation}
One checks easily that $[{\overset{(r)}{\Pi}}]^2 = {\overset{(r)}{\Pi}}, \; r=0,1,2$ and ${\overset{(r)}{\Pi}} {\overset{(s)}{\Pi}} =0$
for $r \neq s$.

Interesting higher-dimensional and complex extensions of Minkowskian space-time were investigated in \cite{Finkelstein}, \cite{BrodyHughston},
albeit without introducing the $Z_3$ grading.

\section{The $Z_3$-graded Lorentz group}

The quadratic Minkowskian square of the $4$ vector $k^{\mu}$, $(k^0)^2 - {\bf k}^2$ is invariant under the transformations
of the Lorentz group. The space rotations touching only the $3$-dimensional vector ${\bf k}$ leave all the three quadratic expressions invariant,
because they depend only on its $3$-dimensional Euclidean square ${\bf k}^2$; therefore we can fix our attention at the Lorentzian
boosts. As we can always align the relative velocity along one of the orthonormal axes of the chosen inertial frame, say $0x$,
those boosts can be considered only between the time and the $x$ coordinates. Here are the three $2 \times 2$ matrices representing
the same Lorentz boost (with real parameter $u$ equal to $\tanh \frac{v}{c}$) leaving invariant one of the three quadratic
invariants given in (\ref{threeinvs}):
\begin{equation}
{\overset{(0) \; \; \; }{L_{00}}} = \begin{pmatrix} \cosh u & \sinh u \cr \sinh u & \cosh u \end{pmatrix}, \; \; \; 
{\overset{(0) \; \; \; }{L_{11}}} = \begin{pmatrix} \cosh u & j^2 \sinh u \cr j \sinh u & \cosh u \end{pmatrix}, \; \; \; 
{\overset{(0) \; \; \; }{L_{22}}} = \begin{pmatrix} \cosh u & j \sinh u \cr j^2 \sinh u & \cosh u \end{pmatrix}, \; \; \; 
\label{Lzero}
\end{equation}
The three matrices are self-adjoint:
\begin{equation}
{\overset{(0) \dagger \; \; }{L_{00}}} = {\overset{(0) \; \; \; }{L_{00}}}, \; \; \; \; 
{\overset{(0) \dagger \; \; }{L_{11}}} = {\overset{(0) \; \; \; }{L_{11}}}, \; \; \; \; 
{\overset{(0) \dagger \; \; }{L_{22}}} = {\overset{(0) \; \; \; }{L_{22}}}, \; \; \; \; 
\label{Lzeroself}
\end{equation}
The above matrices transform each of the three sectors of the $Z_3$-Minkowski space into itself, which founds its reflection 
in the lower indices is quite transparent: $L_{00}$ transforms a vector belonging to the $0$-th sector of the
$Z_3$-graded Minkowskian space into a $4$-vector belonging to the same sector, and similarly for the matrix operators
 $L_{11}$ and $L_{22}$.

It is also easy to prove that each set is a representation of a one-parameter subgroup representing a particular Lorentz
boost, here between the time variable (hereafter always represented by $\tau = ct$) and one cartesian coordinate, say $x$. 
For example, the product of two Lorentz boosts acting on the sector $(1)$, is a boost of the same type:
\begin{equation}
{\overset{(0) \; \; }{L_{11}}} (u) \cdot {\overset{(0) \; \; }{L_{11}}} (v) = {\overset{(0) \; \; }{L_{11}}} (u+v),
\label{groupprop1}
\end{equation}
and similarly for a product of two boosts acting on the sector $(2)$,
\begin{equation}
{\overset{(0) \; \; }{L_{22}}} (u) \cdot {\overset{(0) \; \; }{L_{22}}} (v) = {\overset{(0) \; \; }{L_{22}}} (u+v),
\label{groupprop2}
\end{equation}
The full set of three independent ``classical" (i.e. belonging to the subgroup denoted by ${\overset{(0) \; }{L_{00}}}$) Lorentz boosts
 is given by three $4 \times 4$ matrices, with independent parameters $u, v, w$:
\begin{equation}
\begin{pmatrix} \cosh u & \sinh u & 0 & 0 \cr \sinh u & \cosh u & 0 & 0 \cr 0 & 0 & 1 & 0 \cr 0 & 0 & 0 & 1 \end{pmatrix}, \; \; 
\begin{pmatrix} \cosh v & 0 & \sinh v & 0 \cr 0 & 1 & 0 & 0 \cr \sinh v & 0 & \cosh v & 0 \cr 0 & 0 & 0 & 1 \end{pmatrix}, \; \; 
\begin{pmatrix} \cosh w & 0 & 0 & \sinh w \cr 0 & 1 & 0 & 0  \cr 0 & 0 & 1 & 0 \cr \sinh w & 0 & 0 & \cosh w \end{pmatrix}
\label{threeboosts}
\end{equation} 
To make the extension of the Lorentz boosts complete we need also two sets
of complementary matrix operators transforming one sector into another. There are two types of such operators,
one raising the $Z_3$ index of each subspace, another lowering the $Z_3$ index by $1$. It is quite easy to find out their
matrix representation. 

The matrices lowering the $Z_3$ index by $1$ are::
\begin{equation}
{\overset{(1) \; \; \; }{L_{01}}} = \begin{pmatrix} j \cosh u &  \sinh u \cr j \sinh u & \cosh u \end{pmatrix}, \; \; \; 
{\overset{(1) \; \; \; }{L_{12}}} = \begin{pmatrix} j \cosh u & j^2 \sinh u \cr j^2 \sinh u & \cosh u \end{pmatrix}, \; \; \; 
{\overset{(1) \; \; \; }{L_{20}}} = \begin{pmatrix} j \cosh u & j \sinh u \cr \sinh u & \cosh u \end{pmatrix}, \; \; \; 
\label{Ltwo}
\end{equation}
The determinant of each of these matrices is equal to $j$.
The matrices raising the $Z_3$ index by one (or decreasing it by $2$, which is equivalent from the point of view of the
$Z_3$-grading) are:
\begin{equation}
{\overset{(2) \; \; \; }{L_{10}}} = \begin{pmatrix} j^2 \cosh u & j^2 \sinh u \cr \sinh u & \cosh u \end{pmatrix}, \; \; \; 
{\overset{(2) \; \; \; }{L_{21}}} = \begin{pmatrix} j^2 \cosh u & j \sinh u \cr j \sinh u & \cosh u \end{pmatrix}, \; \; \; 
{\overset{(2) \; \; \; }{L_{02}}} = \begin{pmatrix} j^2 \cosh u & \sinh u \cr j^2 \sinh u & \cosh u \end{pmatrix}, \; \; \; 
\label{Lone}
\end{equation}
The determinant of each of these matrices is equal to $j^2$.
The above sets of three matrices each, decreasing and raising the $Z_3$ index, are mutually hermitian adjoint:
\begin{equation}
{\overset{(1) \dagger \; \; }{L_{01}}} = {\overset{(2) \; \; \; }{L_{10}}}, \; \; \; \; 
{\overset{(1) \dagger \; \; }{L_{12}}} = {\overset{(2) \; \; \; }{L_{21}}}, \; \; \; \; 
{\overset{(1) \dagger \; \; }{L_{20}}} = {\overset{(2) \; \; \; }{L_{02}}}, \; \; \; \; 
\label{Lonetwo}
\end{equation}
Here again, the logic of the lower indices is quite transparent: a matrix labeled $L_{12}$ transforms a $4$-vector belonging to the
sector $(2)$ into a $4$-vector belonging to the sector $(1)$, and so forth, e.g.:
\begin{equation}
L_{01} \; {\overset{(1) \; \; }{k^{\mu}}} = {\overset{(0) \; \; }{k^{\mu'}}}, \; \; \; \; \; 
L_{20} \; {\overset{(0) \; \; }{k^{\mu}}} = {\overset{(2) \; \; }{k^{\mu'}}}, \; \; \; \; \; 
L_{12} \; {\overset{(2) \; \; }{k^{\mu}}} = {\overset{(1) \; \; }{k^{\mu'}}}, \; \; \; \; {\rm etc.}
\label{indicesL}
\end{equation}
The matrices raising or lowering the $Z_3$-grade of the particular type of the $4$-vector they are acting on do not form a group,
because most of the products of two such matrices produce  new matrices not belonging to the set defined above. However, inside
each of one-parameter families corresponding to a given choice of the single space direction concerned by the Lorentz boost,
$0x, \; 0y$ or $0z$ displays the group property if the products are taken according to the chain rule, with second index of the
first factor equal to the first index of the second factor, like in the following examples:
\begin{equation}
{\overset{(1) \; \; \; }{L_{12}}} (\tau, x; \; u)  {\overset{(1) \; \; \; }{L_{20}}} (\tau, x; \; v) =  
{\overset{(2) \; \; \; }{L_{10}}} (\tau, x; \; (u+v)),
\; \; \; 
{\overset{(2) \; \; \; }{L_{21}}} (\tau, y; \; u)  {\overset{(1) \; \; \; }{L_{12}}} (\tau, y; \; v) =  
{\overset{(0) \; \; \; }{L_{22}}} (\tau, y; \; (u+v)),
\;  {\rm etc.}
\label{Lambdaexamples}
\end{equation}

The above $2 \times 2$ matrices represent a reduced version of Lorentz boosts with relative velocity aligned
on the unique axis $Ox$. As in the previous case, the full $4 \times 4$ versions are given by the following three
matrices corresponding to the three independent Lorentz boosts. The boosts of the increasing type, transforming $4$-vectors
from sector $2$ to $0$, from sector $1 $ to $2$ and from sector $0$ to $1$, respectively, are as follows:

- the three boosts ${\overset{(1) \; \; \; }{L_{20}}} (\tau,x),  \; {\overset{(1) \; \; \; }{L_{20}}} (\tau, y), \; 
{\overset{(1) \; \; \; }{L_{20}}} (\tau, z)$ are given by: 
\begin{equation}
\begin{pmatrix} j \cosh u & j \sinh u & 0 & 0 \cr \sinh u & \cosh u & 0 & 0 \cr 0 & 0 & 1 & 0 \cr 0 & 0 & 0 & 1 \end{pmatrix}, \; \; 
\begin{pmatrix} j \cosh v & 0 & j \sinh v & 0 \cr 0 & 1 & 0 & 0 \cr \sinh v & 0 & \cosh v & 0 \cr 0 & 0 & 0 & 1 \end{pmatrix}, \; \; 
\begin{pmatrix} j \cosh w & 0 & 0 & j \sinh w \cr 0 & 1 & 0 & 0  \cr 0 & 0 & 1 & 0 \cr \sinh w & 0 & 0 & \cosh w \end{pmatrix}
\label{threeboosts120}
\end{equation} 
- the three boosts ${\overset{(1) \; \; \; }{L_{12}}} (\tau,x),  \; {\overset{(1) \; \; \; }{L_{12}}} (\tau, y), \; 
{\overset{(1) \; \; \; }{L_{12}}} (\tau, z)$ are given by: 
\begin{equation}
\begin{pmatrix} j \cosh u & j^2 \sinh u & 0 & 0 \cr j^2 \sinh u & \cosh u & 0 & 0 \cr 0 & 0 & 1 & 0 \cr 0 & 0 & 0 & 1 \end{pmatrix}, \; \; 
\begin{pmatrix} j \cosh v & 0 & j^2 \sinh v & 0 \cr 0 & 1 & 0 & 0 \cr j^2 \sinh v & 0 & \cosh v & 0 \cr 0 & 0 & 0 & 1 \end{pmatrix}, \; \; 
\begin{pmatrix} j \cosh w & 0 & 0 & j^2 \sinh w \cr 0 & 1 & 0 & 0  \cr 0 & 0 & 1 & 0 \cr j^2 \sinh w & 0 & 0 & \cosh w \end{pmatrix}
\label{threeboosts112}
\end{equation} 
and the three boosts ${\overset{(1) \; \; \; }{L_{01}}} (\tau,x),  \; {\overset{(1) \; \; \; }{L_{01}}} (\tau, y), \; 
{\overset{(1) \; \; \; }{L_{01}}} (\tau, z)$ are given by: 
\begin{equation}
\begin{pmatrix} j \cosh u & \sinh u & 0 & 0 \cr j \sinh u & \cosh u & 0 & 0 \cr 0 & 0 & 1 & 0 \cr 0 & 0 & 0 & 1 \end{pmatrix}, \; \; 
\begin{pmatrix} j \cosh v & 0 & \sinh v & 0 \cr 0 & 1 & 0 & 0 \cr j \sinh v & 0 & \cosh v & 0 \cr 0 & 0 & 0 & 1 \end{pmatrix}, \; \; 
\begin{pmatrix} j \cosh w & 0 & 0 & \sinh w \cr 0 & 1 & 0 & 0  \cr 0 & 0 & 1 & 0 \cr j \sinh w & 0 & 0 & \cosh w \end{pmatrix}
\label{threeboosts101}
\end{equation} 

The boosts of the decreasing type, transforming $4$-vectors
from sector $1$ to $0$, from sector $2 $ to $1$ and from sector $0$ to $2$, respectively, are as follows:

- the three boosts ${\overset{(2) \; \; \; }{L_{10}}} (\tau,x),  \; {\overset{(2) \; \; \; }{L_{10}}} (\tau, y), \; 
{\overset{(2) \; \; \; }{L_{10}}} (\tau, z)$ are given by: 
\begin{equation}
\begin{pmatrix} j^2 \cosh u & j^2 \sinh u & 0 & 0 \cr \sinh u & \cosh u & 0 & 0 \cr 0 & 0 & 1 & 0 \cr 0 & 0 & 0 & 1 \end{pmatrix}, \; \; 
\begin{pmatrix} j^2 \cosh v & 0 & j^2 \sinh v & 0 \cr 0 & 1 & 0 & 0 \cr \sinh v & 0 & \cosh v & 0 \cr 0 & 0 & 0 & 1 \end{pmatrix}, \; \; 
\begin{pmatrix} j^2 \cosh w & 0 & 0 & j^2 \sinh w \cr 0 & 1 & 0 & 0  \cr 0 & 0 & 1 & 0 \cr \sinh w & 0 & 0 & \cosh w \end{pmatrix}
\label{threeboosts220}
\end{equation} 
- the three boosts ${\overset{(2) \; \; \; }{L_{21}}} (\tau,x),  \; {\overset{(2) \; \; \; }{L_{21}}} (\tau, y), \; 
{\overset{(2) \; \; \; }{L_{21}}} (\tau, z)$ are given by: 
\begin{equation}
\begin{pmatrix} j^2 \cosh u & j \sinh u & 0 & 0 \cr j \sinh u & \cosh u & 0 & 0 \cr 0 & 0 & 1 & 0 \cr 0 & 0 & 0 & 1 \end{pmatrix}, \; \; 
\begin{pmatrix} j^2 \cosh v & 0 & j \sinh v & 0 \cr 0 & 1 & 0 & 0 \cr j \sinh v & 0 & \cosh v & 0 \cr 0 & 0 & 0 & 1 \end{pmatrix}, \; \; 
\begin{pmatrix} j^2 \cosh w & 0 & 0 & j \sinh w \cr 0 & 1 & 0 & 0  \cr 0 & 0 & 1 & 0 \cr j \sinh w & 0 & 0 & \cosh w \end{pmatrix}
\label{threeboosts212}
\end{equation} 
and the three boosts ${\overset{(2) \; \; \; }{L_{02}}} (\tau,x),  \; {\overset{(2) \; \; \; }{L_{02}}} (\tau, y), \; 
{\overset{(2) \; \; \; }{L_{02}}} (\tau, z)$ are given by: 
\begin{equation}
\begin{pmatrix} j^2 \cosh u & \sinh u & 0 & 0 \cr j^2 \sinh u & \cosh u & 0 & 0 \cr 0 & 0 & 1 & 0 \cr 0 & 0 & 0 & 1 \end{pmatrix}, \; \; 
\begin{pmatrix} j^2 \cosh v & 0 & \sinh v & 0 \cr 0 & 1 & 0 & 0 \cr j^2 \sinh v & 0 & \cosh v & 0 \cr 0 & 0 & 0 & 1 \end{pmatrix}, \; \; 
\begin{pmatrix} j^2 \cosh w & 0 & 0 & \sinh w \cr 0 & 1 & 0 & 0  \cr 0 & 0 & 1 & 0 \cr j^2 \sinh w & 0 & 0 & \cosh w \end{pmatrix}
\label{threeboosts202}
\end{equation} 

The nine $4 \times 4$ matrices ${\overset{(r) \; \; \; }{L_{st}}}$, $r, s, t = 0,1,2$ act on the $Z_3$-extended Minkowskian vector in 
a specifically ordered way. Let us write a $Z_3$-extended vector as a column with $12$ entries, composed of three $4$-vectors
belonging each to one of the $Z_3$-graded sectors:
\begin{equation}
( {\overset{(0) \; }{k^{\mu}}}, \;  {\overset{(1) \; }{k^{\mu}}}, \; {\overset{(2) \; }{k^{\mu}}} ) 
\label{triplekvector}
\end{equation}    
\begin{equation}
{\overset{(0) }{\Lambda}} = \begin{pmatrix} {\overset{(0) \; \; \; }{L_{00}}} & 0 & 0 \cr 0 & {\overset{(0) \; \; \; }{L_{11}}} & 0 \cr 
0 & 0 & {\overset{(0) \; \; \; }{L_{22}}} \end{pmatrix} \; \; \; \; \;
{\overset{(1)}{\Lambda}} = \begin{pmatrix} 0 & {\overset{(1) \; \; \; }{L_{01}}} & 0  \cr 0 & 0 &  {\overset{(1) \; \; \; }{L_{12}}} \cr 
{\overset{(1) \; \; \; }{L_{20}}} & 0 & 0 \end{pmatrix} \; \; \; \; \;
{\overset{(2)}{\Lambda}} = \begin{pmatrix} 0 & 0 & {\overset{(2) \; \; \; }{L_{02}}} \cr {\overset{(2) \; \; \; }{L_{10}}} & 0 & 0 \cr 
0 & {\overset{(2) \; \; \; }{L_{21}}} & 0 \end{pmatrix} \; \; \; \; 
\label{Lambda123}
\end{equation}
It is easy to see that the so defined matrices display not only the group property, but also the $Z_3$ grading in the following sense:
\begin{equation}
{\overset{(0) }{\Lambda}}\cdot  {\overset{(0) }{\Lambda}} \subset {\overset{(0) }{\Lambda}}, \; \; \; \;
{\overset{(0) }{\Lambda}}\cdot  {\overset{(1) }{\Lambda}} \subset {\overset{(1) }{\Lambda}}, \; \; \; \; 
{\overset{(0) }{\Lambda}}\cdot  {\overset{(2) }{\Lambda}} \subset {\overset{(2) }{\Lambda}}, \; \; \; \; 
{\overset{(1) }{\Lambda}}\cdot  {\overset{(1) }{\Lambda}} \subset {\overset{(2) }{\Lambda}}, \; \; \; \; 
{\overset{(2) }{\Lambda}}\cdot  {\overset{(2) }{\Lambda}} \subset {\overset{(1) }{\Lambda}}, \; \; \; \; 
{\overset{(1) }{\Lambda}}\cdot  {\overset{(2) }{\Lambda}} = 
{\overset{(2) }{\Lambda}}\cdot  {\overset{(1) }{\Lambda}} \subset {\overset{(0) }{\Lambda}}.
\label{Z3Lambda}
\end{equation}
In other words, the elements of three subsets of the $Z_3$-graded group of boosts behave under associative matrix multiplication 
as folows: 
\begin{equation}
{\overset{(r) }{\Lambda}}\cdot  {\overset{(s) }{\Lambda}} \subset {\overset{(r+s) \mid_3 }{\Lambda}}, \; \; {\rm with} \; \; r, s,.. = 0,1,2, \; \; 
(r+s) \mid_3 = (r+s) \; {\rm modulo} \; \; 3.
\label{Multgrade}
\end{equation}
The three sets of matrices ordered in the particular blocks (\ref{Lambda123}) form a three-parameter family which can be considered
as the extension of the set of three independent Lorentz boosts. In order to obtain the extension of the entire Lorentz group
including the $3$-parameter subgroup of space rotations we shall first investigate the $Z_3$-graded infinitesimal
generators of the Lorentz boosts, and then, taking their commutators, define the $Z_3$-graded extension of the space rotations.

\section{The $Z_3$-graded Lorentz algebra}

The $Z_3$-graded matrix Lie algebra corresponding to the $Z_3$-graded Lie group defined above is easily obtained by
taking the differentials of corresponding families of generators of $1$-parameter abelian subgroups in the vicinity of
the unit element (corfresponding to the $0$ value of the parameter $u$, $v$, etc.). It is sufficient
 to develop all terms in a Taylor series of powers of the parameter $u$ and keep only linear terms
 in the formulae for the matrices of the Lie group defined above, which means that the terms like $1$ or $\cosh u$
will be suppressed, and the terms with $\sinh u$ will be replaced by $1$. Thus, we define: 
\indent
- the full set of three independent ``classical" generators  (i.e. belonging to the subgroup  ${\overset{(0) \; }{\Lambda}}$ 
acting in the first sector of the $Z_3$-extended Mikowski space and which we shall denote 
${\overset{(0) \; }{K_{00}}} (\tau,x),  \; {\overset{(0) \; }{K_{00}}} (\tau, y)$ and ${\overset{(0) \; }{K_{00}}} (\tau,z)$,
the three $4 \times 4$ matrices, defining the boosts between the variables $(\tau, x), \; (\tau, y)$ and $(\tau, z)$, respectively:
acting in the first sector of the $Z_3$-extended Mikowski space
\begin{equation}
{\overset{(0) \; }{K_{00}}} (\tau,x) =
\begin{pmatrix} 0 & 1 & 0 & 0 \cr 1 & 0 & 0 & 0 \cr 0 & 0 & 0 & 0 \cr 0 & 0 & 0 & 0 \end{pmatrix}, \; \; \; \; 
{\overset{(0) \; }{K_{00}}} (\tau, y) =
\begin{pmatrix} 0 & 0 & 1 & 0 \cr 0 & 0 & 0 & 0 \cr 1 & 0 & 0 & 0 \cr 0 & 0 & 0 & 0 \end{pmatrix}, \; \; \; \; 
{\overset{(0) \; }{K_{00}}} (\tau,z) =
\begin{pmatrix} 0 & 0 & 0 & 1 \cr 0 & 0 & 0 & 0  \cr 0 & 0 & 0 & 0 \cr 1 & 0 & 0 & 0 \end{pmatrix}
\label{3K00boosts}
\end{equation} 
as well as two similar sets of $4 \times 4$ matrices, denoted respectively  ${\overset{(0) \; }{K_{11}}}$
and ${\overset{(0) \; }{K_{22}}}$, acting in the sectors $1$ and $2$ of the $Z_3$-extended Minkowskian space
transforming them onto themselves:
\begin{equation}
{\overset{(0) \; }{K_{11}}} (\tau,x) =
\begin{pmatrix} 0 & j^2 & 0 & 0 \cr j & 0 & 0 & 0 \cr 0 & 0 & 0 & 0 \cr 0 & 0 & 0 & 0 \end{pmatrix}, \; \; \; \; 
{\overset{(0) \; }{K_{11}}} (\tau, y) =
\begin{pmatrix} 0 & 0 & j^2 & 0 \cr 0 & 0 & 0 & 0 \cr j & 0 & 0 & 0 \cr 0 & 0 & 0 & 0 \end{pmatrix}, \; \; \; \; 
{\overset{(0) \; }{K_{11}}} (\tau,z) =
\begin{pmatrix} 0 & 0 & 0 & j^2 \cr 0 & 0 & 0 & 0  \cr 0 & 0 & 0 & 0 \cr j & 0 & 0 & 0 \end{pmatrix}
\label{3K11boosts}
\end{equation} 
transforming the sector $1$ onto itself, and
\begin{equation}
{\overset{(0) \; }{K_{22}}} (\tau,x) =
\begin{pmatrix} 0 & j & 0 & 0 \cr j^2 & 0 & 0 & 0 \cr 0 & 0 & 0 & 0 \cr 0 & 0 & 0 & 0 \end{pmatrix}, \; \; \; \; 
{\overset{(0) \; }{K_{22}}} (\tau, y) =
\begin{pmatrix} 0 & 0 & j & 0 \cr 0 & 0 & 0 & 0 \cr j^2 & 0 & 0 & 0 \cr 0 & 0 & 0 & 0 \end{pmatrix}, \; \; \; \; 
{\overset{(0) \; }{K_{22}}} (\tau,z) =
\begin{pmatrix} 0 & 0 & 0 & j \cr 0 & 0 & 0 & 0  \cr 0 & 0 & 0 & 0 \cr j^2 & 0 & 0 & 0 \end{pmatrix}
\label{3K22boosts}
\end{equation} 
transforming sector $2$ onto itself.

There are also the two sets
of complementary matrix operators transforming sectors into one another. There are two types of such infinitesimal generators,
one raising the $Z_3$ index of each subspace, another decreasing the $Z_3$ index by $1$. Their matrix representation is as follows:
\vskip 0.2cm
\indent
- the infinitesimal generators of three boosts ${\overset{(1) \; \; \; }{K_{20}}} (\tau,x),  \; {\overset{(1) \; \; \; }{K_{20}}} (\tau, y), \; 
{\overset{(1) \; \; \; }{K_{20}}} (\tau, z)$ are given by: 
\begin{equation}
{\overset{(1) \; \; \; }{K_{20}}} (\tau,x) =
\begin{pmatrix} 0 & j  & 0 & 0 \cr 1 & 0 & 0 & 0 \cr 0 & 0 & 0 & 0 \cr 0 & 0 & 0 & 0 \end{pmatrix}, \; \; \; \; 
{\overset{(1) \; \; \; }{K_{20}}} (\tau, y) =
\begin{pmatrix} 0 & 0 & j  & 0 \cr 0 & 0 & 0 & 0 \cr 1 & 0 & 0 & 0 \cr 0 & 0 & 0 & 0 \end{pmatrix}, \; \; \; \;
{\overset{(1) \; \; \; }{K_{20}}} (\tau, z) =
\begin{pmatrix} 0 & 0 & 0 & j  \cr 0 & 0 & 0 & 0  \cr 0 & 0 & 0 & 0 \cr 1 & 0 & 0 & 0 \end{pmatrix}
\label{Algboosts120}
\end{equation} 
- the three infinitesimal generators of boosts ${\overset{(1) \; \; \; }{K_{12}}} (\tau,x),  \; {\overset{(1) \; \; \; }{K_{12}}} (\tau, y), \; 
{\overset{(1) \; \; \; }{K_{12}}} (\tau, z)$ are given by: 
\begin{equation}
{\overset{(1) \; \; \; }{K_{12}}} (\tau, x) =
\begin{pmatrix} 0 & j^2 & 0 & 0 \cr j^2 & 0 & 0 & 0 \cr 0 & 0 & 0 & 0 \cr 0 & 0 & 0 & 0 \end{pmatrix}, \; \; \; \; 
{\overset{(1) \; \; \; }{K_{12}}} (\tau, y) =
\begin{pmatrix} 0 & 0 & j^2 & 0 \cr 0 & 0 & 0 & 0 \cr j^2  & 0 & 0 & 0 \cr 0 & 0 & 0 & 0 \end{pmatrix}, \; \; \; \;
{\overset{(1) \; \; \; }{K_{12}}} (\tau, z) =
\begin{pmatrix} 0 & 0 & 0 & j^2 \cr 0 & 0 & 0 & 0  \cr 0 & 0 & 0 & 0 \cr j^2 & 0 & 0 & 0 \end{pmatrix}
\label{Algboosts112}
\end{equation} 
and the three boosts ${\overset{(1) \; \; \; }{K_{01}}} (\tau,x),  \; {\overset{(1) \; \; \; }{K_{01}}} (\tau, y), \; 
{\overset{(1) \; \; \; }{K_{01}}} (\tau, z)$ are given by: 
\begin{equation}
{\overset{(1) \; \; \; }{K_{01}}} (\tau, x) =
\begin{pmatrix} 0 & 1 & 0 & 0 \cr j  & 0 & 0 & 0 \cr 0 & 0 & 0 & 0 \cr 0 & 0 & 0 & 0 \end{pmatrix}, \; \; \; \; 
{\overset{(1) \; \; \; }{K_{01}}} (\tau, y) =
\begin{pmatrix} 0 & 0 & 1 & 0 \cr 0 & 0 & 0 & 0 \cr j  & 0 & 0 & 0 \cr 0 & 0 & 0 & 0 \end{pmatrix}, \; \; \; \;
{\overset{(1) \; \; \; }{K_{01}}} (\tau, z) =
\begin{pmatrix} 0 & 0 & 0 & 1 \cr 0 & 0 & 0 & 0  \cr 0 & 0 & 0 & 0 \cr j  & 0 & 0 & 0 \end{pmatrix}
\label{Algboosts101}
\end{equation} 

The boosts of the increasing type, transforming $4$-vectors
from sector $0$ to $1$, from sector $1 $ to $2$ and from sector $2$ to $0$, respectively, are as follows:

- the three infinitesimal boosts ${\overset{(2) \; \; \; }{K_{10}}} (\tau,x),  \; {\overset{(2) \; \; \; }{K_{10}}} (\tau, y), \; 
{\overset{(2) \; \; \; }{K_{10}}} (\tau, z)$ are given by: 
\begin{equation}
{\overset{(2) \; \; \; }{K_{10}}} (\tau,x) =
\begin{pmatrix} 0 & j^2 & 0 & 0 \cr 1 & 0 & 0 & 0 \cr 0 & 0 & 0 & 0 \cr 0 & 0 & 0 & 0 \end{pmatrix}, \; \; \; \; 
{\overset{(2) \; \; \; }{K_{10}}} (\tau, y) =
\begin{pmatrix} 0 & 0 & j^2 & 0 \cr 0 & 0 & 0 & 0 \cr 1 & 0 & 0 & 0 \cr 0 & 0 & 0 & 0 \end{pmatrix}, \; \; \; \; 
{\overset{(2) \; \; \; }{K_{10}}} (\tau, z) =
\begin{pmatrix} 0 & 0 & 0 & j^2  \cr 0 & 0 & 0 & 0  \cr 0 & 0 & 0 & 0 \cr 1 & 0 & 0 & 0 \end{pmatrix}
\label{Algboosts220}
\end{equation} 
- the three boosts ${\overset{(2) \; \; \; }{K_{21}}} (\tau,x),  \; {\overset{(2) \; \; \; }{K_{21}}} (\tau, y), \; 
{\overset{(2) \; \; \; }{K_{21}}} (\tau, z)$ are given by: 
\begin{equation}
{\overset{(2) \; \; \; }{K_{21}}} (\tau, x) =
\begin{pmatrix} 0 & j & 0 & 0 \cr j & 0 & 0 & 0 \cr 0 & 0 & 0 & 0 \cr 0 & 0 & 0 & 0 \end{pmatrix}, \; \; \; \; 
{\overset{(2) \; \; \; }{K_{21}}} (\tau, y) =
\begin{pmatrix} 0 & 0 & j & 0 \cr 0 & 0 & 0 & 0 \cr j & 0 & 0 & 0 \cr 0 & 0 & 0 & 0 \end{pmatrix}, \; \; \; \;
{\overset{(2) \; \; \; }{K_{21}}} (\tau, z) =
\begin{pmatrix} 0 & 0 & 0 & j \cr 0 & 0 & 0 & 0  \cr 0 & 0 & 0 & 0 \cr j  & 0 & 0 & 0 \end{pmatrix}
\label{Algboosts212}
\end{equation} 
and the three boosts ${\overset{(2) \; \; \; }{K_{02}}} (\tau,x),  \; {\overset{(2) \; \; \; }{K_{02}}} (\tau, y), \; 
{\overset{(2) \; \; \; }{K_{02}}} (\tau, z)$ are given by: 
\begin{equation}
{\overset{(2) \; \; \; }{K_{02}}} (\tau, x) =
\begin{pmatrix} 0 & 1 & 0 & 0 \cr j^2  & 0 & 0 & 0 \cr 0 & 0 & 0 & 0 \cr 0 & 0 & 0 & 0 \end{pmatrix}, \; \; \; \; 
{\overset{(2) \; \; \; }{K_{02}}} (\tau, y) =
\begin{pmatrix} 0 & 0 & 1 & 0 \cr 0 & 0 & 0 & 0 \cr j^2  & 0 & 0 & 0 \cr 0 & 0 & 0 & 0 \end{pmatrix}, \; \; \; \;
{\overset{(2) \; \; \; }{K_{02}}} (\tau, z) =
\begin{pmatrix} 0 & 0 & 0 & 1 \cr 0 & 0 & 0 & 0  \cr 0 & 0 & 0 & 0 \cr j^2 & 0 & 0 & 0 \end{pmatrix}
\label{Algboosts202}
\end{equation} 
The so defined infinitesimal generators keep the symmetry properties of the Lie group matrices, i.e. they close under the commutator
product, provided that the two factors satisfy the chain rule, with the second index of the first matrix coinciding with the
first index of the second matrix, like in the following examples:
\begin{equation}
\left[ {\overset{(2) \; \; \; }{K_{02}}} (\tau, y), {\overset{(2) \; \; \; }{K_{21}}} (\tau, z) \right], \; \; \; 
\left[ {\overset{(1) \; \; \; }{K_{20}}} (\tau, z), {\overset{(2) \; \; \; }{K_{02}}} (\tau, x) \right], \; \; {\rm etc.}
\label{Kcommutators}
\end{equation}

The $27$ generators form three groups containing three matrices each, belonging to raising, lowering or neutral type with
respect to the $Z_3$-grade of the Minkowskian $4$-vector

\begin{equation}
{\overset{(0) }{K}} = \begin{pmatrix} {\overset{(0) \; \; \; }{K_{00}}} & 0 & 0 \cr 0 & {\overset{(0) \; \; \; }{K_{11}}} & 0 \cr 
0 & 0 & {\overset{(0) \; \; \; }{K_{22}}} \end{pmatrix} \; \; \; \; 
{\overset{(1)}{K}} = \begin{pmatrix} 0 & {\overset{(1) \; \; \; }{K_{01}}} & 0  \cr 0 & 0 &  {\overset{(1) \; \; \; }{K_{12}}} \cr 
{\overset{(1) \; \; \; }{K_{20}}} & 0 & 0 \end{pmatrix} \; \; \; \; 
{\overset{(2)}{K}} = \begin{pmatrix} 0 & 0 & {\overset{(2) \; \; \; }{K_{02}}} \cr {\overset{(2) \; \; \; }{K_{10}}} & 0 & 0 \cr 
0 & {\overset{(2) \; \; \; }{K_{21}}} & 0 \end{pmatrix} \; \; \; \; 
\label{AlgK123}
\end{equation}
%
Each of the three big $12 \times 12$ matrices composed of three blocks of $4 \times 4$ matrices ${\overset{(p) \; \; \; }{K_{rs}}}$
($p,r,s = 0,1,2$) appears in three different versions corresponding to the choice of one of the three elementary Lorentz boosts
in $(\tau, x), \; (\tau, y)$ or $(\tau, z)$ $2$-dimensional spacetime planes. Let us denote them by ${\overset{(r) }{{\cal{K}}_i}}, \; i=1,2,3$,
corresponding to the respective choice of the space direction $x, y$ or $z$. For example, for ${\overset{(1) }{{\cal{K}}_y}}$ we shall
get explicitly
\begin{equation}
{\overset{(1) }{{\cal{K}}_y}} = \begin{pmatrix} 0 & {\overset{(1) \; \; \; }{K_{01}}} (\tau,y) & 0  
\cr 0 & 0 &  {\overset{(1) \; \; \; }{K_{12}}} (\tau, y) \cr 
{\overset{(1) \; \; \; }{K_{20}}} (\tau, y) & 0 & 0 \end{pmatrix},
\label{ExKy}
\end{equation} 
and so forth.

The spatial rotations around the axes $0x, \; 0y$ and $0z$ are represented in the usual $4$-dimensional Minkowskian space as follows:
\begin{equation}
J_x = \begin{pmatrix} 0 & 0 & 0 & 0 \cr 0 & 0 & 0 & 0 \cr 0 & 0 & 0 & -1 \cr 0 & 0 & 1 & 0 \end{pmatrix}, \; \; \; 
J_y = \begin{pmatrix} 0 & 0 & 0 & 0 \cr 0 & 0 & 0 & 1 \cr 0 & 0 & 0 & 0 \cr 0 & -1 & 0 & 0 \end{pmatrix}, \; \; \; 
J_z = \begin{pmatrix} 0 & 0 & 0 & 0 \cr 0 & 0 & -1 & 0 \cr 0 & 1 & 0 & 0 \cr 0 & 0 & 0 & 0 \end{pmatrix}. 
\label{ThreeJ}
\end{equation}
The full set of $12 \times 12$ matrices representing three independent spatial rotations acting on the
twelve-dimensional $Z_3$-graded Minkowskian spacetime is as follows:
\begin{equation}
{\overset{(0)}{\cal{J}}}_i = \begin{pmatrix} J_i & 0 & 0 \cr 0 & J_i & 0 \cr 0 & 0 & J_i \end{pmatrix}, \; \; \; \; \; \; 
{\overset{(1)}{\cal{J}}}_i = \begin{pmatrix} 0 & J_i & 0 \cr 0 & 0 & J_i \cr J_i & 0 & 0 \end{pmatrix}, \; \; \; \; \; \; 
{\overset{(2)}{\cal{J}}}_i = \begin{pmatrix} 0 & 0 & J_i \cr J_i & 0 & 0 \cr 0 & J_i & 0 \end{pmatrix}, \; \; \; \; \; \; 
\label{ThreespaceJ}
\end{equation}
They also form a $Z_3$ graded Lie algebra with respect to the ordinary Lie bracket (the commutator of matrices).  
Therefore we get the full set of $Z_3$-graded relations defining the algebra ($r, s, \; r+s$ are modulo $3$), conformally with the structure
of the $Z_3$-graded Lorentz algebra introduced in \cite{RKJL2019}.
\begin{equation}
[ \; {\overset{(r) }{{\cal{K}}_i}}, \; {\overset{(s) }{{\cal{K}}_k}} \; ]  = - \epsilon_{ikl} {\overset{(r+s) }{{\cal{J}}_l}}, \; \; \; \; \; 
[ \; {\overset{(r) }{{\cal{J}}_i}}, \; {\overset{(s) }{{\cal{K}}_k}} \; ]  = \epsilon_{ikl} {\overset{(r+s) }{{\cal{K}}_l}}, \; \; \; \; \;%
[ \; {\overset{(r) }{{\cal{J}}_i}}, \; {\overset{(s) }{{\cal{J}}_k}} \; ] = \epsilon_{ikl} {\overset{(r+s) }{{\cal{J}}_l}}.
\label{modulocomm}
\end{equation}

\section{$Z_3$-extended Poincar\'e algebra and the Casimir operators}

The standard Poincar\'e algebra is the semi-direct product of the Lorentz algebra and the $4$-dimensional abelian algebra of
translations $P_{\mu}$, satisfying the well-known commutation relations:

$$ \left[ M_{\mu \nu}, M_{\lambda \rho} \right] = \eta_{\mu \rho} M_{\nu \lambda} -  \eta_{\nu \lambda} M_{\mu \rho}
+ \eta_{\mu \lambda} M_{\nu \rho} -  \eta_{\nu \rho} M_{\mu \lambda},$$
\begin{equation}
\left[ P_{\mu} , P_{\nu} \right] = 0, \; \; \; \; \; \; \; 
\left[ M_{\mu \nu}, P_{\lambda} \right] =  \eta_{\mu \lambda} P_{\nu} - \eta_{\nu \lambda} P_{\mu},
\label{Poincarealg}
\end{equation}
In terms of six generators $K_i = M_{0i}$ and $J_m = \frac{1}{2} \; \epsilon_{ikm} M_{ik}$, $i,k,.. = 1,2,3$, the standard commutation relations
\begin{equation}
\left[ J_i, J_k \right] = \epsilon_{ikl} J_l, \; \; \; \; \; \left[ J_i, K_k \right] = \epsilon_{ikl} K_l, 
\; \; \; \; \; \; 	
\left[ K_i, K_k \right] = - \epsilon_{ikl} J_l.
\label{modulocomm2}
\end{equation}
must be complemented by the following extra commutation relations with $P_{\mu} = (P_0, P_i)$: 
\begin{equation}
\left[ K_i, P_0 \right] = P_i, \; \; \; \; \left[K_i , P_j \right] = - \delta_{ij} \; P_0, \; \; \; \; \; 
\left[ J_i , P_0 \right] = 0, \; \; \; \; \; \left[ J_i , P_k \right] = \epsilon_{ikm} P_m.
\label{extracom}
\end{equation}
The most appropriate realization of the totality of commutation relations given by (\ref{modulocomm}) and (\ref{extracom})
is via differential operators, with the generators $P_{\mu}$ identified with partial derivations $\partial_{\mu}$. These operators
can be produced from the standard matrix representation by the following well-known procedure. Let us take for example the
$4 \times 4$ matrix representation of $3$-dimensional rotations given by formulae (\ref{ThreeJ}). The differential operators
corresponding to $J_x, \; J_y$ and $J_z$ are obtained by taking formally the scalar product of the space-time $4$-covector
$[\tau, x, y, z]$ with the $4$-gradient $\partial_{\mu}$ transformed by the corresponding matrix $J_i$. Take for example the
matrix $J_x$:
\begin{equation} 
\begin{pmatrix} \tau, & x, & y, & z \end{pmatrix} 
\begin{pmatrix} 0 & 0 & 0 & 0 \cr 0 & 0 & 0 & 0 \cr 0 & 0 & 0 & -1 \cr 0 & 0 & 1 & 0 \end{pmatrix}
\begin{pmatrix} \partial_{\tau} \cr \partial_x \cr \partial_y \cr \partial_z \end{pmatrix} = z \partial_y - y \partial_z.
\label{Jxdiff}  
\end{equation}
Similarly we get $ J_y \rightarrow x \partial_z - z \partial_x, \; \; \; \; \; J_z \rightarrow y \partial_x - x \partial_y.$

Our next aim is to extend the standard Poincar\'e algebra so as to include the $Z_3$-graded Lorentz algebra defined by
the set of commutation relations (\ref{modulocomm}) complemented by the set of three types of translation generators,
denoted by ${\overset{(r) }{{\cal{P}}_{\mu}}}$, $r = 0,1,2$ and $\mu, \nu,..=0,1,2,3$. Let us separate time and space components;
we shall write then
\begin{equation}
{\overset{(r) }{{\cal{P}}_{\mu}}} = [ {\overset{(r) }{{\cal{P}}_{0}}}, {\overset{(r) }{{\cal{P}}_{i}}} ].
\label{P0Pi}
\end{equation}
We expect the following $Z_3$-graded generalization of standard commutation relations between the Lorentz and translation generators:
\begin{equation}
[ \; {\overset{(r) }{{\cal{P}}_0}}, \; {\overset{(s) }{{\cal{P}}_k}} \; ]  = 0; \; \; \; \; \;
[ \; {\overset{(r) }{{\cal{P}}_i}}, \; {\overset{(s) }{{\cal{P}}_j}} \; ]  = 0,
\label{PPcomm}
\end{equation}
\begin{equation}
[ \; {\overset{(r) }{{\cal{J}}_k}}, \; {\overset{(s) }{{\cal{P}}_0}} \; ]  = 0; \; \; \; \; \; 
[ \; {\overset{(r) }{{\cal{J}}_i}}, \; {\overset{(s) }{{\cal{P}}_k}} \; ] = \epsilon_{ikl}\; {\overset{(r+s) }{{\cal{P}}_l}}, 
\label{JPgradcomm}
\end{equation}
\begin{equation}
[ \; {\overset{(r) }{{\cal{K}}_i}}, \; {\overset{(s) }{{\cal{P}}_0}} \; ]  =  {\overset{(r+s) }{{\cal{P}}_i}}, \; \; \; \; \; 
[ \; {\overset{(r) }{{\cal{K}}_i}}, \; {\overset{(s) }{{\cal{P}}_k}} \; ]  = - \delta_{ik}  {\overset{(r+s) }{{\cal{P}}_0}}.
\label{KJPextended}
\end{equation}
In all the above relations the grades $r, s = 0,1,2$ add up modulo $3$.

The construction of differential operators providing faithful representation of the $Z_3$-graded Poincar\'e algebra (\ref{KJPextended})
we shall follow the prescription given by (\ref{Jxdiff}) with $12 \times 12$ matrices introduced in previous section, and $12$-component
generalizations of Minkowskian $4$-vectors and co-vectors. Let us introduce the following notation for generalized vectors in triple
Minkowskian space-time:
\begin{equation}
\left[ \tau_0, x_0, y_0, z_0; \; \tau_1, x_1, y_1, z_1; \; \tau_2, x_2, y_2, z_2 \right],
\label{12vector}
\end{equation}
The notations are obvious: the lower index ``$0$" refers to the standard Minkowskian component (graded $0$), while the indices ``$1$" 
and ``$2$" refer to two complex extensions, mutually conjugate, of $Z_3$ grades $1$ and $2$, respectively.

For the moment we leave aside the definition of metrics in the so extended triple Minkowskian space-time.


Partial derivatives take, with respect to these variables are represented by the following $12$-component column vector 
(written here as a horizontal co-vector transposed, in order to spare the space):
\begin{equation}
\left[ \; \partial_{\tau_0}, \partial_{x_0}, \partial_{y_0}, \partial_{z_0}; \;   
\partial_{\tau_1}, \partial_{x_1}, \partial_{y_1}, \partial_{z_1}; \;
\partial_{\tau_2}, \partial_{x_2}, \partial_{y_2}, \partial_{z_2}; \;   \right]^T
\label{12gradient}
\end{equation}
What is left now is to compute patiently the results of contraction of the co-vector (\ref{12vector}) with the $12$-component
generator of generalized translations (\ref{12gradient}) with one of the eighteen $12 \times 12$ matrices representing the
 generalized Lorentz algebra (\ref{modulocomm}) sandwiched in between. This will produce the $18$ generators of the
$Z_3$-graded Poincar\'e algebra represented in form of linear differential operators. With twelve translations (\ref{P0Pi})
we shall get the $30$-dimensional $Z_3$-graded extension of the Poincar\'e algebra, of which the usual $10$-dimensional
subalgebra is the standard Poincar\'e algebra.

The results are a bit cumbersome, but their construction and symmetry properties are quite clear.

Let us start with the nine generalized Lorentz boosts ${\overset{(r)}{\cal{K}}}_i$. We have explicitly:

$${\overset{(0)}{\cal{K}}_x} = (\tau_0 \partial_{x_0} + x_0 \partial_{\tau_0}) + ( j^2 \; \tau_1 \partial_{x_1} +j \; x_1 \partial_{\tau_1} )
+ ( j \; \tau_2 \partial_{x_2} + j^2 \; x_2 \partial_{\tau_2} ), $$
$${\overset{(0)}{\cal{K}}_y} = (\tau_0 \partial_{y_0} + y_0 \partial_{\tau_0}) + ( j^2 \; \tau_1 \partial_{y_1} +j \; y_1 \partial_{\tau_1} )
+ ( j \; \tau_2 \partial_{y_2} + j^2 \; y_2 \partial_{\tau_2} ), $$
\begin{equation}
{\overset{(0)}{\cal{K}}_z} = (\tau_0 \partial_{z_0} + z_0 \partial_{\tau_0}) + ( j^2 \; \tau_1 \partial_{z_1} +j \; z_1 \partial_{\tau_1} )
+ ( j \; \tau_2 \partial_{z_2} + j^2 \; z_2 \partial_{\tau_2} ); 
\label{Kzero}
\end{equation}
$${\overset{(1)}{\cal{K}}_x} = (\tau_0 \partial_{x_1} + j \; x_0 \partial_{\tau_1}) + ( j^2 \; \tau_1 \partial_{x_2} + j^2 \; x_1 \partial_{\tau_2} )
+ ( j \; \tau_2 \partial_{x_0} +   x_2 \partial_{\tau_0} ), $$
$${\overset{(1)}{\cal{K}}_y} = (\tau_0 \partial_{y_1} + j \; y_0 \partial_{\tau_1}) + ( j^2 \; \tau_1 \partial_{y_2} + j^2 \; y_1 \partial_{\tau_2} )
+ ( j \; \tau_2 \partial_{y_0} +  y_2 \partial_{\tau_0} ), $$
\begin{equation}
{\overset{(1)}{\cal{K}}_z} = (\tau_0 \partial_{z_1} + j \; z_0 \partial_{\tau_1}) + ( j^2 \; \tau_1 \partial_{z_2} + j^2 \; z_1 \partial_{\tau_2} )
+ ( j \; \tau_2 \partial_{z_0} +  z_2 \partial_{\tau_0} ); 
\label{Kuno}
\end{equation}
$${\overset{(2)}{\cal{K}}_x} = (\tau_0 \partial_{x_2} + j^2 \; x_0 \partial_{\tau_2}) + ( j \; \tau_2 \partial_{x_1} + j \; x_2 \partial_{\tau_1} )
+ ( j^2 \; \tau_1 \partial_{x_0} + x_1 \partial_{\tau_0} ), $$
$${\overset{(2)}{\cal{K}}_y} = (\tau_0 \partial_{y_2} + j^2 \; y_0 \partial_{\tau_2}) + ( j \; \tau_2 \partial_{y_1} + j \; y_2 \partial_{\tau_1} ) 
+ ( j^2 \; \tau_1 \partial_{y_0} + y_1 \partial_{\tau_0} ), $$
\begin{equation}
{\overset{(2)}{\cal{K}}_z} = (\tau_0 \partial_{z_2} + j^2 \; z_0 \partial_{\tau_2}) + ( j \; \tau_2 \partial_{z_1} + j \; z_2 \partial_{\tau_1} ) 
+ ( j^2 \; \tau_1 \partial_{z_0} + z_1 \partial_{\tau_0} ).    .
\label{Kdos}
\end{equation}
The $Z_3$-graded generalized differential operators representing the Lorentz boosts display remarkable symmetry properties.
The ``diagonal" generators ${\overset{(0)}{\cal{K}}_i}$ are hermitian: they are invariant under the simultaneous complex
conjugation, replacing $j$ by $j^2$ and vice versa, and switching the indices $1 \rightarrow 2, \; 2 \rightarrow 1$.

Under the same hermitian symmetry operation the $Z_3$-graded boosts ${\overset{(1)}{\cal{K}}_i}$ and ${\overset{(2)}{\cal{K}}_i}$
transform into each other, so that we have 
$${\overset{(1)\dagger}{{\cal{K}}_i}} = {\overset{(2)}{\cal{K}}_i}, \; \; \; \; 
{\overset{(2)\dagger}{{\cal{K}}_i}} = {\overset{(1)}{\cal{K}}_i}.$$

The commutation relations between the generalized Lorentz boosts given by (\ref{Kzero}, \ref{Kuno}) and (\ref{Kdos})
define the differential representation of $Z_3$-graded extension of pure rotations, ${\overset{(r)}{{\cal{J}}_k}}$,
with $r= 0,1,2$ and $i,j,..= 1,2,3$. By tedious (but not too sophisticated) calculation we can check that the
commutation relations between the $Z_3$-graded Lorentz boosts imposed as hypothesis in (\ref{modulocomm}):
$$[ \; {\overset{(r) }{{\cal{K}}_i}}, \; {\overset{(s) }{{\cal{K}}_k}} \; ]  = - \epsilon_{ikl} {\overset{(r+s) }{{\cal{J}}_l}},$$  
lead indeed to the following expressions for spatial rotations ${\overset{(s)}{{\cal{J}}_i}}$:
$$ {\overset{(0)}{{\cal{J}}_x}} = (z_0 \partial_{y_0} - y_0 \partial_{z_0}) + (z_1 \partial_{y_1} - y_1 \partial_{z_1}) + 
(z_2 \partial_{y_2} - y_2 \partial_{z_2}),$$
 $$ {\overset{(0)}{{\cal{J}}_y}} = (x_0 \partial_{z_0} - z_0 \partial_{x_0}) + (x_1 \partial_{z_1} - z_1 \partial_{x_1}) + 
(x_2 \partial_{z_2} - z_2 \partial_{x_2}),$$
\begin{equation} 
{\overset{(0)}{{\cal{J}}_z}} = (y_0 \partial_{x_0} - x_0 \partial_{y_0}) + (y_1 \partial_{x_1} - x_1 \partial_{y_1}) + 
(y_2 \partial_{x_2} - x_2 \partial_{y_2}),
\label{J0k}
\end{equation}
Note that the above generators are sums of classical expressions for $J_k$, each of them acting in its own sector of the $Z_3$-graded
extension of Minkowskian space-time. 

The grade $1$ generators of rotations ${\overset{(1)}{{\cal{J}}_i}}$ have the same form, but mix up coordinates with derivatives 
from different sectors, in cyclical order, symbolically $0 \rightarrow 1, \; 1 \rightarrow 2, \; 2 \rightarrow 0$:

$$ {\overset{(1)}{{\cal{J}}_x}} = (z_0 \partial_{y_1} - y_0 \partial_{z_1}) + (z_1 \partial_{y_2} - y_1 \partial_{z_2}) + 
(z_2 \partial_{y_0} - y_2 \partial_{z_0}),$$
 $$ {\overset{(1)}{{\cal{J}}_y}} = (x_0 \partial_{z_1} - z_0 \partial_{x_1}) + (x_1 \partial_{z_2} - z_1 \partial_{x_2}) + 
(x_2 \partial_{z_0} - z_2 \partial_{x_0}),$$
\begin{equation} 
{\overset{(1)}{{\cal{J}}_z}} = (y_0 \partial_{x_1} - x_0 \partial_{y_1}) + (y_1 \partial_{x_2} - x_1 \partial_{y_2}) + 
(y_2 \partial_{x_0} - x_2 \partial_{y_0}),
\label{J1k}
\end{equation}
Finally, the grade $2$ generators of spatial rotations, ${\overset{(2)}{{\cal{J}}_i}}$, repeat the same scheme, but in reverse (anti-cyclic) order,
i.e. $0 \rightarrow 2, \; 1 \rightarrow 0, \; 2 \rightarrow 1$:
$$ {\overset{(2)}{{\cal{J}}_x}} = (z_0 \partial_{y_2} - y_0 \partial_{z_2}) + (z_1 \partial_{y_0} - y_1 \partial_{z_0}) + 
(z_2 \partial_{y_1} - y_2 \partial_{z_1}),$$
 $$ {\overset{(2)}{{\cal{J}}_y}} = (x_0 \partial_{z_2} - z_0 \partial_{x_2}) + (x_1 \partial_{z_0} - z_1 \partial_{x_0}) + 
(x_2 \partial_{z_1} - z_2 \partial_{x_1}),$$
\begin{equation} 
{\overset{(2)}{{\cal{J}}_z}} = (y_0 \partial_{x_2} - x_0 \partial_{y_2}) + (y_1 \partial_{x_0} - x_1 \partial_{y_0}) + 
(y_2 \partial_{x_1} - x_2 \partial_{y_1}),
\label{J2k}
\end{equation}
It can easily be checked that these differential operators correspond to what we would get by direct construction using the
matrix representation given in (\ref{ThreespaceJ}). The $18$ differential operators acting on the $Z_3$-graded extension of
Minkowskian space-time; the $9$ generalized Lorentz boosts ${\overset{(r)}{{\cal{K}}_i}}$ and the $9$ generalized space
rotations ${\overset{(s)}{{\cal{J}}_k}}$, with $\; r, s = 0,1,2$ and $i,j = 1,2,3$, define the faithful representation
of the $Z_3$-graded generalization of the Lorentz group.

In order to introduce the extension to full Poincar\'e group we have to add three $4$-component generators of translations
each one acting on its own sector of the generalized $Z_3$-graded Minkowskian space-time. It turns out that in order to
satisfy the $Z_3$-graded set of standard commutation relations given by (\ref{KJPextended}), the three differential operators
 $$ {\overset{(0) }{{\cal{P}}_{\mu}}}, \; \; \; \;  {\overset{(1) }{{\cal{P}}_{\mu}}}, \; \; \; \; {\overset{(2) }{{\cal{P}}_{\mu}}} $$
must be defined as follows:
\begin{equation}
{\overset{(0) }{{\cal{P}}_{\mu}}} = \left[ \; \partial_{\tau_0}, \; - \partial_{x_0}, \; - \partial_{y_0}, \; - \partial_{z_0} \; \right]
\label{Pzero}
\end{equation}
\begin{equation}
{\overset{(1) }{{\cal{P}}_{\mu}}} = \left[ \; j \partial_{\tau_1}, \; - \partial_{x_1}, \; - \partial_{y_1}, \; - \partial_{z_1} \; \right]
\label{Puno}
\end{equation}
\begin{equation}
{\overset{(2) }{{\cal{P}}_{\mu}}} = \left[ \; j^2  \partial_{\tau_2}, \; - \partial_{x_2}, \; - \partial_{y_2}, \; - \partial_{z_2} \; \right]
\label{Pdos}
\end{equation}

It can be checked by direct computation that the eighteen  generators $ {\overset{(r)}{{\cal{K}}_i}}$ and  $ {\overset{(s)}{{\cal{J}}_k}}$
together with the twelve generalized $Z_3$-graded translations defined above by (\ref{Pzero}, \ref{Puno}, \ref{Pdos}) satisfy the
full set of $Z_3$-graded extension of the Poincar\'e algebra. Its total dimension is $30$, three times ten, corresponding to three
replicas of the classical Poincar\'e group, one ``diagonal", acting on three components of the $Z_3$-graded Minkowskian space-time
 without mixing them, and two other replicas acting on all three components transforming them into one another. The commutations 
relations are given by the set defined in (\ref{P0Pi}, \ref{PPcomm}, \ref{JPgradcomm}) and (\ref{KJPextended}).

Classical Poincar\'e algebra admits two Casimir operators which commute with all generators. These are the $4$-square of the
translation $4$-vector $P_{\mu} P^{\mu}$, and the $4$-square of the Pauli-Lubanski $4$-vector $W_{\mu} W^{\mu}$, where
\begin{equation}
W^{\mu} = \frac{1}{2} \; \varepsilon^{\mu \nu \lambda \rho} \; J_{\nu \lambda} P_{\rho}, \; \; \; \; 
J_{0 i} = K_i, \; \; \; J_{ik} = \epsilon_{ikl} J^{l}. 
\label{PauliLubanski}
\end{equation}  
In terms of more familiar generators $K_i$ and $J_l$ the Pauli-Lubanski vector takes on the following form:
\begin{equation}
W_0 = J_i P^i = {\bf J} \cdot {\bf P}, \; \; \; \;  \; \; W_i = P_0 \; J_i - \epsilon_{ijk} P^j K^k, \; \; {\rm or}
\; \; {\bf W} = P^0 \; {\bf J} - {\bf P} \wedge {\bf K}.
\label{Wvect}
\end{equation} 
The following relations are easily verified:
\begin{equation}
W_{\mu} P^{\mu} = 0, \; \; \; \; \left[ W^{\mu}, P^{\lambda} \right] = 0, \; \; \; \; 
\left[ J^{\mu \lambda} , W^{\rho} \right] = \eta^{\lambda \rho} W^{\mu} - \eta^{\mu \rho} W^{\lambda}.
\label{WPrelations}
\end{equation}
 
The eigenvalues of these two Casimir operators, corresponding to the mass and orbital spin of a given particle state,
define the irreducible representations of the Poincar\'e group,
\begin{equation}
P_{\mu} P^{\mu} = m^2, \; \; \; \; W_{\mu} W^{\mu} = L(L+1)
\label{masspin}
\end{equation} 
In the case of the $Z_3$-graded extension the corresponding Casimir operators must be invariant under permutations
imposed by the $Z_3$ symmetry. That is to say, the three types of generators should contribute equally to the
generalized Casimir operator. The expression generalizing the mass operator $P_{\mu} P^{\mu}$ should contain
not only the obvious term ${\overset{(0)}{{\cal{P}}_{\mu}}} {\overset{(0)}{{\cal{P}}^{\mu}}}$, but also other
contributions of all possible grades, like e.g. another grade $0$ term: 
${\overset{(1)}{{\cal{P}}_{\mu}}} {\overset{(2)}{{\cal{P}}^{\mu}}}$, as well as other similar terms of grades $1$ and $2$. 
The symmetric and {\it real} combination imitating the first Casimir operator in (\ref{masspin}) is as follows:
\begin{equation}
{\cal{P}}^2 =  {\overset{(0)}{{\cal{P}}_{\mu}}} {\overset{(0)}{{\cal{P}}^{\mu}}} + {\overset{(1)}{{\cal{P}}_{\mu}}} {\overset{(1)}{{\cal{P}}^{\mu}}}
+ {\overset{(2)}{{\cal{P}}_{\mu}}} {\overset{(2)}{{\cal{P}}^{\mu}}} + {\overset{(0)}{{\cal{P}}_{\mu}}} {\overset{(1)}{{\cal{P}}^{\mu}}}
+ {\overset{(1)}{{\cal{P}}_{\mu}}} {\overset{(2)}{{\cal{P}}^{\mu}}} + {\overset{(2)}{{\cal{P}}_{\mu}}} {\overset{(0)}{{\cal{P}}^{\mu}}},
\label{Z3PCasimir}
\end{equation}
The Pauli-Lubanski $4$-vector also possesses its $Z_3$-graded extensions. They are of the following form:
$${\overset{(0)}{{\cal{W}}_{\mu}}} = \frac{1}{2} \varepsilon_{\mu \nu \lambda \rho} 
({\overset{(0)}{{\cal{J}}^{\nu \lambda} }} {\overset{(0)}{{\cal{P}}^{\rho}}} + 
{\overset{(1)}{{\cal{J}}^{\nu \lambda} }} {\overset{(2)}{{\cal{P}}^{\rho}}} +
{\overset{(2)}{{\cal{J}}^{\nu \lambda} }} {\overset{(1)}{{\cal{P}}^{\rho}}} ), $$
$${\overset{(1)}{{\cal{W}}_{\mu}}} = \frac{1}{2} \varepsilon_{\mu \nu \lambda \rho} 
({\overset{(2)}{{\cal{J}}^{\nu \lambda} }} {\overset{(2)}{{\cal{P}}^{\rho}}} + 
{\overset{(1)}{{\cal{J}}^{\nu \lambda} }} {\overset{(0)}{{\cal{P}}^{\rho}}} +
{\overset{(0)}{{\cal{J}}^{\nu \lambda} }} {\overset{(1)}{{\cal{P}}^{\rho}}} ), $$
\begin{equation}
{\overset{(2)}{{\cal{W}}_{\mu}}} = \frac{1}{2} \varepsilon_{\mu \nu \lambda \rho} 
({\overset{(1)}{{\cal{J}}^{\nu \lambda} }} {\overset{(1)}{{\cal{P}}^{\rho}}} + 
{\overset{(2)}{{\cal{J}}^{\nu \lambda} }} {\overset{(0)}{{\cal{P}}^{\rho}}} +
{\overset{(0)}{{\cal{J}}^{\nu \lambda} }} {\overset{(2)}{{\cal{P}}^{\rho}}} ).
\label{ThreeW}
\end{equation}
With these three graded Pauli-Lubanski vectors we can produce a $Z_3$-invariant extended Casimir operator
of orbital spin:
\begin{equation}
{\cal{W}}^2 = {\overset{(0)}{{\cal{W}}_{\mu}}} {\overset{(0)}{{\cal{W}}^{\mu}}} + {\overset{(1)}{{\cal{W}}_{\mu}}} {\overset{(1)}{{\cal{W}}^{\mu}}}
+ {\overset{(2)}{{\cal{W}}_{\mu}}} {\overset{(2)}{{\cal{W}}^{\mu}}} + {\overset{(0)}{{\cal{W}}_{\mu}}} {\overset{(1)}{{\cal{W}}^{\mu}}}
+ {\overset{(1)}{{\cal{W}}_{\mu}}} {\overset{(2)}{{\cal{W}}^{\mu}}} + {\overset{(2)}{{\cal{W}}_{\mu}}} {\overset{(0)}{{\cal{W}}^{\mu}}},
\label{Z3WCasimir}
\end{equation}

The analysis of eigenvalues of the generalized Casimir operators and the classification of irreducible representations
of $Z_3$-graded extension of the Poincar\'e algebra presented here will be the subject of the forthcoming publications.

%
%
%

\vskip 0.5cm
\indent
{\large{\bf Acknowledgement}}
\vskip 0.3cm
\indent
\hskip 0.5cm
The author is greatly indebted to Jerzy Lukierski for countless discussions, enlightening remarks and lots of very useful suggestions.

\end{document}